\documentclass[journal]{IEEEtran}
\usepackage{cite}
\usepackage{amsmath,amssymb,amsfonts}
\usepackage{algorithmic}
\usepackage{graphicx}
\usepackage{textcomp}
\usepackage{xcolor}
\usepackage{amsmath,amsfonts}
\usepackage{algorithmic}
\usepackage{algorithm}
\usepackage{array}
\usepackage[caption=false,font=normalsize,labelfont=sf,textfont=sf]{subfig}
\usepackage{textcomp}
\usepackage{stfloats}
\usepackage{url}
\usepackage{verbatim}
\usepackage{graphicx}
\usepackage{cite}
\usepackage{hhline}
\usepackage{tabularx}
\usepackage{booktabs}
\usepackage{color,soul}
\setulcolor{red}
\sethlcolor{yellow}
\usepackage{xcolor}
\usepackage{amsthm}
\usepackage[english]{babel}

\usepackage{listings}
\usepackage{color}
\renewcommand\hl[1]{#1} 
\begin{document}
%
\title{In-Orbit Optical SSA Using Proliferated LEO Satellites for Space Traffic Monitoring: An Analytical Framework}
%
%
%

\author{Dianle~Gong~and Peng~Hu,~\IEEEmembership{Senior Member,~IEEE}
\thanks{D. Gong is with Department of Electrical and Computer Engineering, University of Manitoba, Winnipeg, Canada R3T 2N2, and with the Department of Applied Mathematics, University of Waterloo, Ontario, Canada N2L 3G1}
\thanks{P. Hu is with Department of Electrical and Computer Engineering, University of Manitoba, Winnipeg, Canada R3T 2N2. Corresponding author: Peng Hu, E-mail: peng.hu@umanitoba.ca}
}

%
%

\markboth{Journal of \LaTeX\ Class Files,~Vol.~00, No.~0, 00~2026}%
{Shell \MakeLowercase{\textit{et al.}}: Bare Demo of IEEEtran.cls for IEEE Journals}
%



\maketitle

\begin{abstract}
The increase in space activities has increased the risks of space debris generation, affecting space safety and sustainability. Traditional space situational awareness (SSA) relies on single star trackers and ground-based tracking facilities. There is limited discussion on the use of in-orbit optical sensors on low Earth orbit (LEO) satellite constellations for SSA, despite their importance for efficient space traffic management systems. In this paper, we aim to address this important challenge. We first present a new analytical system model for utilizing LEO satellite constellations for in-orbit SSA. We then develop a method to evaluate and analyze such a system. We also propose a Poisson expected revisit period algorithm and introduce the period of equivalent orbital distributions to reveal the relationship between revisit period and geometric variables, with insightful results based on real-world and custom satellite constellations. Experiments on real-world constellation show that the representative Poisson expected revisit period ranges from 0.4 days to 5.7 days for targets whose apogee altitude ranges from 552 km to 650 km, while requiring a per-case computation time of 0.4 s to 4.8 s. Our work can inform the future design of in-orbit and onboard computing systems for SSA, such as space object detection and space traffic monitoring systems.
\end{abstract}


%
\IEEEpeerreviewmaketitle

\begin{table}
\caption{Nomenclature}
\label{Paper_Nomenclature}
\resizebox{\columnwidth}{!}{
\begin{tabular}{ll}
\toprule
$F$ & Field of view \\
$U$ &The target being monitored\\
$w$ & \hl{Work distance} \\
$\gamma,\delta$ & Pixel dimensions of the rectangular CCD sensor array. \\
& We choose $\gamma=\delta$ here.\\
$r$ & Current distance from the center of the Earth\\
$\hat{r}$ & Radial directional vector\\
$\vec{v_r}$ & Relative velocity\\
$u_1$ & Relative velocity component perpendicular to the sensor line of sight\\
$u_2$ & Relative velocity component parallel to the sensor line of sight\\
$v_0$ & Constellation orbital speed\\
$\vec{v_1}$ & Velocity vector of $U$\\ 
$\vec{v_2}$ & Projected velocity vector of a satellite\\
$\vec{v_1'}$ & Transverse component of $\vec{v_1}$\\
$\vec{v_2'}$ & Transverse component of $\vec{v_2}$\\
$n_v$ & Dimensionless speed of $U$\\
$m_v$ & Projected dimensionless speed of a satellite\\
$\kappa$ & Least number of pixels for detection\\
$l$ & Diameter of the target\\
$G$ & Gravity constant\\
$C_a$ & Constellation coverage of availability\\
$C_s$ & Stationary coverage\\
$r_1$ & Inner radius of the monitored shell\\
$r_2$ & Outer radius of the monitored shell\\
$h$ & altitude\\
$h_0$ &Nominal altitude of the satellite constellation\\
$h_1$ &Nominal altitude of the target when the target is on a circular orbit\\
$T$ & Poisson expected revisit period\\
$T_0$ & Constellation orbit period\\
$T_{0U}$ & Target orbit period\\
$T_r$ &Real revisit time\\
$T_1$ &Period of orbital beat\\
$T_2$ &Period of equivalent orbital distribution\\
$T_3$ &Period of scanning\\
$\theta$ & True anomaly of the target on its elliptical orbit\\
$\phi$ & Orbital phase angle\\
$\varphi_{or}$ & Relative right ascension of ascending node (RAAN) phase \\
& between the target orbit and the constellation\\
$\rho$ & Radius of the Earth\\
$\rho_i$ & Radius of different layers of constellations\\
$\rho_{air}(h)$ & Density of Air at altitude $h$\\
$a,b,c,e$ & Semi-major Axis, Semi-minor Axis, Semi-focal length and linear eccentricity\\

$P$ & Probability of detecting the target during a single orbit alignment\\
$\lambda$ & Mean generation rate in a Poisson process\\

$i$ & Walker-delta constellation orbit inclinations\\
$i_U$ & Target orbit inclinations\\
$t$ & Total number of the satellites in the constellation\\
$p$ & Total number of the orbit planes in the constellation\\
$f$ & Relative phase between adjacent orbits\\
$\alpha$ & Pointing angle\\
PA & Perigee altitude\\
AA & Apogee altitude\\
$\omega$ & Angular velocity\\
$\omega_U$ & Angular velocity of $U$\\
$\Delta\omega$ & Precessional angular velocity\\
$\Delta\omega$ & J2 precessional angular velocity \\
$\vec{j}$& Directional vector within the pyramid\\
$\vec{e}$& Directional vector fo the line of relative motion\\
$\Delta r$& Altitude difference determined by $h_1-h_0$\\
$\Delta r'$ & y-axis coordinate in the pyramid-$U$ collision problem\\
$\tau,\tau'$ & Distance along the lines determined by the directional vectors $\vec{e}$ and $\vec{j}$\\
$d$ & Distance between adjacent satellites in the same orbital plane\\
$K.E.$ & Kinetic energy\\
$F_k$ & The $k$th force\\
$\vec{r_k}$ & The $k$th position vector\\
$E$ & Orbital energy\\
$B$ & ballistic coefficient\\
$A$ & Aerodynamical area\\
$M, m, m_U$ & Mass of the Earth, mass of the satellite, and mass of $U$\\
$mu$ & Gravitational parameter equals to  $GM$\\
\bottomrule
\end{tabular}
}
\end{table}

\section{Introduction}
The growth of space activities, represented by the expansion of low Earth orbit (LEO) satellites, have led to a significant growth of anthropogenic objects in near-Earth space, including satellites, rocket bodies, and large amounts of debris. These not only exacerbate the near-Earth space environment but also crowd the space traffic, endangering the safety of spacecraft operations, astronauts, and missions. 

The threats posed by the increasing space debris and close proximity of nano/small satellites in LEO due to intensive space activities are real and expected to be exacerbated in the future based on the space debris report from European Space Agency (ESA) and the well-known Kessler syndrome \cite{kessler_ml} that theoretically indicates chain reactions from from cascading collisions arising from space debris. The recent incidents regarding the risks posed by the close-contact satellites to the saftey operations of the spacecraft in LEO. For example, on May 23, 2022, Starlink 3890 and Starlink 3896 came within 8 meters of each other in orbit. Later, in November 2025, Starlink 35956 experienced a malfunction that led to tumbling and orbital decay, ultimately generating multiple pieces of space debris. On March 29, 2026, Starlink 34343 had an on-orbit anomaly leading to space debris generation. 

Space situational awareness (SSA) has been a general approach to tracking and detecting space objects. Among all the SSA systems, vision-based sensor based constellation systems are recieving attentions and could be an important solution to SSA. This type of system can consist of optical vision sensors that scan the space, capable of detecting and tracking target space objects. The utilization of vision sensors has shown its promises from the space traffic management systems such as the vision-based SpaceX Stargaze system deployed in January 2026 and the efficiency of the use of onboard vision sensors have been evident in \cite{Gaposchkin2000, ZhangHu_2024_WiSEE, ZhangHu_2025_SpaceOps}. However, the understanding of the large deployment for on-orbit spacecraft, such as LEO satellites, and the deployment principles, mechanisms, and dynamics of these systems are lacking. 

The works in multiple points of view \cite{Li2020} and a complete analysis on sensor tasking analysis\cite{XUE2024} depict value and efforts in sensor system structuring and emphasized on performance analysis. The conjunction data messages (CDMs) \cite{acciarini-2020-spacecraft} have become a feasible mechanism for detecting and predicting. The generation of CDMs are mostly based on the ground monitoring of space objects where the in-situ and real-time onboard detection mechanism on spacecraft is lacking. However, the approach has shown limitations in recent challenges rising from the intensive space activities. For example, the stringent performance requirements, payload constraints for communication and computing, and the insufficient autonomous operational solutions have hardly been addressed in the literature. To address these emerging challenges, space/satellite object detection (SOD) has recently been proposed \cite{ZhangHu_2024_WiSEE, ZhangHu_2025_SpaceOps, ZhangHu_2025_WiSEE} as a means to enhance the space sustainability and safety issues, with the deep learning (DL) approach showing effective performance in detecting the satellite objects using single and collaborative satellites \cite{ZhangHu_2025_WiSEE} in LEO constellations. SOD has become an enabling approach to efficient AI-based solutions \cite{Hu2023_SatAIOps} for satellite constellations. 

Despite these advances, an important gap remains. Existing studies have shown the promise of space-based optical sensing and vision-based SOD, but they do not provide a dedicated analytical framework for deploying and evaluating onboard optical sensors for close-range observation in LEO. In particular, prior studies either focus on designing optical based LEO satellite system to monitor GEO targets \cite{xie2022} or rely on simulation-based evaluation \cite{acciarini-2020-spacecraft, 2Anniballe2025}, where the critical relationships among sensor geometry, constellation configuration, revisit performance, and orbital decay are insufficiently characterized. 

To fill this gap, this paper develops an analytical model for object detection using optical systems in LEO constellations, introduces revisit-time- and availability-oriented performance metrics, and proposes a Poisson-based framework to quantify detection performance and its evolution under orbital decay. Leveraging its object detection capability, the constellation can form a vision-based space system for SSA and space traffic management (STM), capable of detecting orbiting objects using multiple optical sensors, and cooperate with current ESA models \cite{ESAMOD}.

The main contributions of the paper are listed as follows:
\begin{itemize}
    \item Propose a system model for optoelectronic based SSA systems in LEO.
    \item Develop a methodology to dynamically analyze and evaluate system performance using both classical and newly defined metrics; for example, revisit time and stationary coverage are included as classical metrics, and availability-based coverage as a new metric.
    \item Propose the \textit{Poisson expected revisit period} algorithm to characterize the relationship between revisit periods and geometric variables, enabling analysis of the impact of orbital decay. 
    \item Evaluate and discuss the proposed methods in LEO constellation case studies and provide practical guidelines for developing optical based SSA systems.
\end{itemize}

The remainder of this paper is organized as follows. Section II reviews the related work. Section III presents the problem statement and system model. Section IV describes the proposed algorithm. Section V discusses the numerical results and analytical insights. Section VI presents case studies. Section VII provides concluding remarks and directions for future work.

\begin{figure}[htbp]
\centerline{\includegraphics[width=\columnwidth]{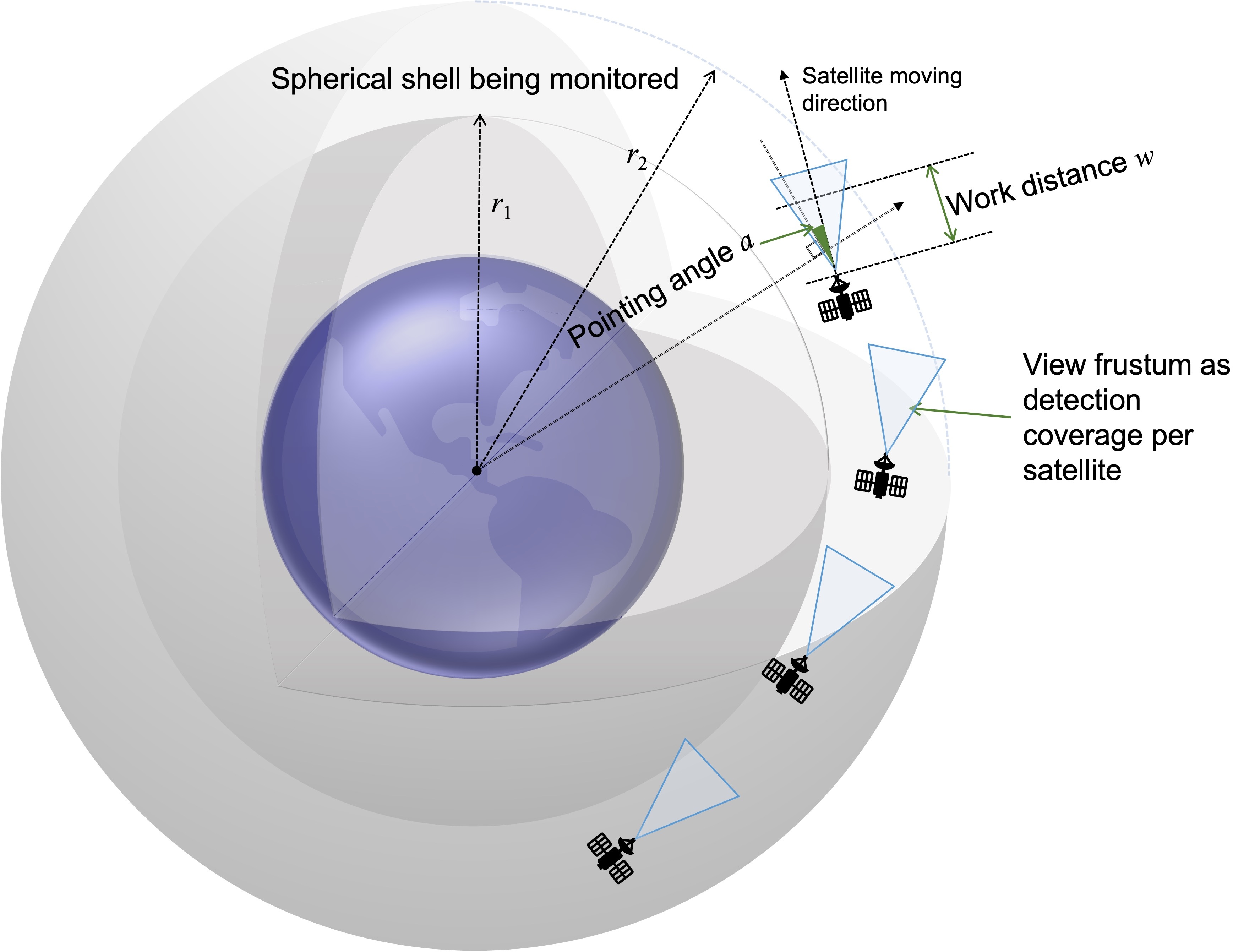}}
\caption{Geometry of the satellite-based optical sensing system in LEO. (The scale of LEOs is illustrative.)}
\label{fig:geometry_overview}
\end{figure}

\section{Related Work}
The rapid growth of space activities has made space debris monitoring increasingly challenging. The monitoring and tracking of spacecraft and objects can be covered in the overall topic of SSA. Within this context, new technologies are specifically required to support timely collision avoidance. A key component of collision avoidance is orbit prediction, where the spacecraft orbits in advance can be computed based on their updated ephemeris. 

Since the early 2000s, the long-term risk of cascading debris-generating collisions, commonly referred to as the Kessler syndrome, has been recognized and continuously researched for decades \cite{IADC2004} due to possibilities of catastrophic consequences. Some real collision incidents then occurred in history. Iridium 33 collided with Kosmos 2251 in 2009. More recently, in 2025, Starlink 35956 experienced an anomaly that led to tumbling, orbital decay, and the creation of additional space debris. From the report by European Space Agency \cite{ESA2025}, there are more than 30,000 cataloged space debris objects larger than 10 cm and the majority of them are located in LEO. 

A critical first step toward reducing such collision occurrences is the development of a space-based surveillance system (SBSS) capable of identifying and tracking debris and satellites in LEO orbits. Studies on space-based optical platforms, including space-based visible (SBV) systems \cite{Gaposchkin2000}, show that deploying visible-spectrum sensors in orbit enhances both coverage and measurement accuracy for geostationary orbit (GEO) and other high-altitude resident space objects (RSOs) relative to purely ground-based approaches. Similarly, Vanwijck and Flohrer \cite{Flohrer2008} indicated space-based sensors can significantly strengthen SSA by capturing observations that are often inaccessible to ground systems, particularly in regions that are difficult to monitor or insufficiently covered. Their work positions space-based optical sensing as a valuable complement to terrestrial surveillance, offering advantages in coverage, revisit frequency, and observation geometry. An early work on SBSS highlighted its distinction from previous constellation designs and introduced the above-the-horizon (ATH) constraint \cite{Takano2012}. However, that work relied on existing star sensors. Since then, advances in sensor technology have significantly enhanced the ability to identify targets at lower cost.

With the advent of machine learning (ML) techniques and reusable launch vehicles, the application of cost-effective optical sensors in SSA has recently begun to be explored. Anniballe \textit{et al.} \cite{1Anniballe2025, 2Anniballe2025} explored and developed the constraints for optical sensors capable of searching for targets thousands of kilometers away, and widely analyzed multiple metrics from coverage to maximum revisit time with Monte Carlo. Also, Xie \textit{et al.} developed a model to deal with the GEO revisit period \cite{xie2022} analytically, though it is designed to be applicable in particular scenarios. However, there is no analysis on small satellites and are not applicable for the situation we are going to discuss.

Some recent work uses a computer vision approach for detecting space objects, including debris, as satellite/space object detection (SOD) tasks have emerged in the literature. Some of the works included multi-point vision, whose performance is more complex than most models can discuss\cite{XUE2024}.
Continuing the work of small CNN optical sensor satellites that could be deployed distributed \cite{ZhangHu_2025_WiSEE}, we propose an analytical method to evaluate any symmetric constellation consisting of large amounts of light optical sensors with unique visibility constraints, with new metrics. Also, we will consider how orbital decay will influence these metrics of the constellation.

\section{Problem Statement \& System Model}

\subsection{Problem Statement}
In the LEO there is a satellite constellation in the Walker-Delta constellation with $p$ orbital planes and a total of $t$ satellites with an inclination of $i$, orbital radius of $\rho+h_0$ where $h_0$ is the nominal altitude of the satellite constellation. Each satellite is equipped with an optical sensor pointing $\alpha$ from its direction of motion as shown in Fig. \ref{fig:geometry_overview}. The optical sensor \cite{optics_basics} has a field of view (FoV) of $F$ radians with a work distance, $w$, also presented in Fig. \ref{fig:geometry_overview}. 

Based on this system setup, we aim to achieve the following objectives:
\begin{itemize}
    \item Find and calculate appropriate metrics such as a real revisit time $T_r$, stationary coverage $C_s$ to evaluate the target satellite optical sensor constellation. Considering the near-Earth spherical shell between the altitude of $r_1$ and $r_2$ ($r_1 < r_2$), the spherical shell region being monitored is defined as
    \begin{equation} 
        A = \{ x \in \mathbb{R}^n \mid  r_1 + \rho < \|x\| < r_2 + \rho \},
    \end{equation}
    
    where $\rho$ is the Earth radius (as a sphere) and thus $r_1 + \rho$ and $r_2 + \rho$ are the inner and outer radii of the two concentric spheres.
    \item  If all or a portion of the satellites in such a constellation experience orbital decay caused by orbital perturbations, how would the major metrics change quantitatively?
\end{itemize}

\subsection{Proposed System Model}

We will describe the proposed system model in the subsequent sub-sections.
\subsubsection{Optical Based Sensors}
An optical based sensor discussed is a square CCD array with pixel dimensions $\gamma\times \delta$, where we assume $\gamma=\delta$ due to clarity without losing generality. This is the linear resolution for any picture. 

All optical based sensors used in satellites point in a direction that is $\alpha$ above its direction of motion, as shown in Fig. \ref{fig:geometry_overview}. When $\alpha=0$, the optical sensor of the satellite is pointing toward the direction of motion all the time. In an example where $l=1$ m, $\gamma=2048$, $F=\frac{\pi}{4}$, the corresponding range of detection is approximately $w=2608$ m. If target identification takes $\kappa$ pixels, then the range will be multiplied by $\frac{1}{\sqrt{\kappa}}$. The value of $w$ means that $C_s$, the stationary coverage for temporary observation coverage, will be continuously too small to provide useful results.

\subsubsection{Constellation}
We consider the common constellations for LEO satellites, which are  Walker-delta and Walker-star. Walker-Delta constellations are defined by a set of data consisting of four parameters $i:t/p/f$, where $i$ is the inclination, $t$ is the total number of satellites, $p$ is the number of equally spaced planes, and $f$ is the relative spacing between satellites in adjacent planes. A Walker-Star constellation with $p$ orbital planes at inclination $i=\frac{\pi}{2}$ is geometrically related to a Walker-Delta constellation with $2p$ planes at the same inclination, where each pair of counter-rotating planes shares the same ground track but carries satellites in opposite directions.

Other constellation configurations, such as the elliptical Walker–Delta constellation, may offer advantages for certain metrics, particularly $C_a$.

\subsubsection{Detection Target Parameters}

Let $U$ denote a recognizable target being monitored under consideration in this study. A recognizable target means $U$ is  an individually discernible object or debris to a LEO satellite, which is essential to fine-grained detectability for SSA and STM. The choice of target here is object-level: metrics that evaluate the SSA system should focus on successive observations of the same target $U$, not merely the repeated coverage of an anonymous orbital region. This principle is emphasized in \cite{Sanchez2013}, and is considered and implemented in several important subfields of SSA, including SST (Space Surveillance and Tracking) \cite{saezbo2018} and SBSS (Space-Based Space Surveillance) \cite{2013ESASP.723E..27E}. 
$U$ is assumed to follow an arbitrary inclined elliptical orbit, which is a common solution to Kepler dynamic equations \cite{goldstein2002} for arbitrary space objects. This orbital configuration is characterized by its inclination $i_U$, the semi-major axis $a$, and the linear eccentricity $c$. Over a sufficiently short time scale, the orbits in $U$ can be regarded as stationary. 
For generality, we also assume $U$ is uniformly distributed in $i_U$, perigee altitude (PA), and apogee altitude (AA). 

An important metric of the SSA  performance is the revisit time. In our model, we consider two important revisit times. $T$ is the Poisson expected revisit time, which is a theoretical value that determines the expected time interval between two observations for uniform Poisson generation rate. However, the assumption of the uniform Poisson generation rate does not hold in real world, because the time intervals between observations are not completely randomized when the orbit periods of both satellites and $U$ are close to each other. According to the beat theory, the Poisson generation rate will oscillate under this circumstance, with a period of time scale $T_2$. For one recognizable target, considering $T_2$, we introduce $T_r$ as the real revisit time, which is the expected time interval between two observations of $U$, 
\begin{equation}
    T_r=Max(T,T_2)
    \label{eq_Tr}
\end{equation}

For multiple recognizable targets, $T_r$ over these targets will be an average value following $\frac{1}{T_r}=\Sigma\frac{1}{T_{ri}} / N_U$ based on the Poisson assumption, where $N_U$ is the number of targets being monitored.


\subsubsection{Availability}

For a given $w$, a Walker–Delta constellation can generally observe only a subset of the near-Earth spherical shell. As a result, some target orbits in $U$ may not be observed. We can define a coverage describing the availability of the constellation optical sensors. This motivates the introduction of a constellation coverage of availability, denoted by $C_a$, to describe the capability.

For $U$ that contain any possible targets in a sphere shell between the radius of $r_1$ and $r_2$, integration over the perigee altitude and AA yields an unnormalized measure proportional to  $(r_1-r_2)^2$, which depends only on the separation between the two shell boundaries. It then follows that $C_a$ for 
$n$ Walker–Delta constellations can be written as
\begin{equation}
C_a=1-\frac{\Sigma (\rho_{i+1}-\rho_{i})^2}{(r_2-r_1)^2}
\label{eq_Ca}
\end{equation}
where $\rho_i$ is the radius of $i$-th layer Walker-delta constellation, $\rho_0=r_1$ and $\rho_{n+1}=r_2$. The optimal case is obtained when the Walker-delta constellations are spaced equally.

\subsubsection{Inhomogeneity}

In this case, one may ask whether there exists one or a set of particular relative orbital configurations and distributions that never enter the FoV of any optical sensor. A representative scenario arises when the targets in $U$ and the sensor constellation occupy similar orbits. In this orbit configuration, the relative phase of the optical sensors and the targets changes very slowly. Generally, the targets are not always in the same orbit as the satellites, where detectable targets are usually located at a different altitude. The resulting small velocity difference between the optical sensors and the targets will form a beat-like relative motion. As a consequence, the relative orbital spatial distribution evolves gradually through phase space and may repeatedly approach certain quasi-stationary traces, along which the targets remain inaccessible to all sensors. These traces are in unavailable bands in the phase space. For a given design parameter $f$, the width of the unavailable bands is inversely related to $\frac{t}{p}$ which is the number of optical sensors in one orbit plane because the optical sensors are unrecognizable. If the beat period $T_1$ is sufficiently small, the phase distribution may be approximated as nearly uniform; otherwise, the revisit behavior is primarily governed by $T_1$. 

Generally, $T_1$ is determined by orbital periods of satellites and $U$, which are $T_0$ and $T_{0U}$, whose relation could be represented by the following equation:
\begin{equation}
    \frac{1}{T_1}=|\frac{1}{T_0}-\frac{1}{T_{0U}}|
    \label{eq_T1'}
\end{equation}

Following the standard two-body Kepler relation \cite{goldstein2002}, $T_0$ and the angular velocity $\omega$ are determined by
\begin{equation}
T_0=2\pi\sqrt{\frac{(\rho+h_0)^3}{GM}}
\label{eq_T0}
\end{equation}
and
\begin{equation}
\omega=\sqrt{\frac{GM}{(\rho+h_0)^3}}.
\label{eq_o}
\end{equation}
Similarly, If $U$ is on a circular orbit whose altitude is $h_1$, then the semi-major axis is $a=\rho+h_1$, and there are $T_{0U}$ and $\omega_U$ determined by
\begin{equation}
T_{0U}=2\pi\sqrt{\frac{a^3}{GM}}
\label{eq_T0}
\end{equation}
and
\begin{equation}
\omega_U=\sqrt{\frac{GM}{a^3}}.
\label{eq_o}
\end{equation}
 According to \eqref{eq_T0}, by differentiating  \eqref{eq_o} we get the precessional angular velocity $\Delta\omega$:
\begin{equation}
     \Delta\omega=-\frac{2(\Delta r)}{3(\rho+h_0)}\omega
\label{eq_do}
\end{equation}
where the altitude difference $\Delta r=h_1-h_0$. 
Therefore, $T_1$ is determined by:
\begin{equation}
T_1=\frac{3(\rho+h_0)}{2(\Delta r)}T_0
\label{eq_T1}
\end{equation}

where $G$ is the gravitational constant $G=6.674\times10^{-11}$, $M$ is the mass of the earth, $\rho$ is the radius of the earth, $h_0$ is the altitude of the sensors,  $h_1$ is the altitude of $U$. 

There are $\frac{p}{t}$ satellites in the same orbit, so the beat phenomena does not have to oscillate a complete beat period, but only a phase change of $\frac{2\pi p}{t}$. Let $T_2$ represent the period of equivalent orbital distributions of both $U$ and the constellation. It is determined by:
\begin{equation}
    T_2=\frac{T_1 p}{t}
    \label{eq_T2}
\end{equation} and if $T_2$ is larger than $T$, then $T_r$ should be adjusted to $T_2$.

 The precession caused by the J2 perturbation is another source of relative phase change. It gives rise to a small precessional angular velocity and is therefore also expected to influence the beat period. However, theoretical estimates indicate that the precessional angular velocity induced by the J2 perturbation is less than $0.1 \%$ of the beat angular speed, suggesting that this effect is negligible. Based on  \cite{strizzi1993improved}, the J2 precessional angular velocity $\Delta \Omega$ is shown in \eqref{eq_dO} and \eqref{eq_dO'}:

\begin{equation}
\Delta\Omega=-\frac{21}{4}J_2\frac{\rho^2}{(\rho+h_0)^\frac{9}{2}}\sqrt{GM}\cos{i}(\Delta r)
\label{eq_dO}
\end{equation}
 
 or 
 \begin{equation}
      \Delta\Omega=-\frac{21}{4}J_2\frac{\rho^2}{(\rho+h_0)^3}\omega\cos{i}(\Delta r)
      \label{eq_dO'}
 \end{equation}
 where $J_2=1.083\times10^{-3}$. Compared with \eqref{eq_o} and \eqref{eq_do}, this suggests that the precessing caused by $J_2$ gravity is ignorable.

\subsubsection{Relative velocities}

For any target $U$, the origin is also the same centre of gravity as the whole constellation. If we pick up a single target and a single satellite, the two orbital planes must intersect along a line passing through the origin. Because of the symmetry, the relative velocities at the two alignment points on the intersection line are centrally symmetric, making it sufficient to evaluate only one alignment point where the $U$ and the satellite are aligned to get all the information.

Assume the constellation orbital speed is $v_0$ at $\rho+h_0$. $v_1$ is the velocity of $U$ and $v_2$ is the projected velocity of a satellite. $v_2$ is obtained by projecting the angular velocity of the satellite to the position of $U$. We can assume $|\vec{v_1}|=n_v v_0$, $|\vec{v_2}|=m_v v_0$ where $n_v,m_v$ are the dimensionless speeds. The relative velocity $\vec{v_r}$ is 
\begin{equation}
\vec{v_r}=\vec{v_1}-\vec{v_2}.
\label{eq_vr'}
\end{equation}
 $n_v$ is the dimensionless speed of $U$:
\begin{equation}
    n_v=\sqrt{\frac{\rho+h_0}{\rho+h_1}}
\label{eq_nv}
\end{equation}
and $m_v$ is the projected dimensionless speed of the satellite:
$
    m_v=\frac{\rho+h_1}{\rho+h_0}
    \label{eq_mv}
$, and $\vec{n_3}$ is the unit vector along the line of intersection. $v_0$ is the linear velocity of the satellite equals to $\sqrt{\frac{GM}{\rho+h_0}}$.

In an ellipse situation, use the transverse velocity as the magnitude in place of $v_0$. 
\begin{equation}
|\vec{v_1'}|=\frac{\sqrt{a^2-c^2}}{r}\sqrt{\frac{GM}{a}}
\label{eq_v1'}
\end{equation}
where $c$ is the semi-focal length, and
\begin{equation}
|\vec{v_2'}|=|\vec{v_2}|=\frac{\rho+h_1}{\rho+h_0}\sqrt{\frac{GM}{\rho+h_0}}
\label{eq_v2'}
\end{equation}
Thus, for low$e$ situations, \eqref{eq_vr'} is now:
\begin{equation}
    \vec{v_r}=(\vec{v_1}'-\vec{v_2}')
\label{eq_vr}
\end{equation}

The speed of the object at a given radial distance can be obtained from the standard two-body Kepler relation, namely the vis-viva equation \cite{goldstein2002},
\begin{equation}
    \frac{1}{2}mv^2-\frac{GMm}{r}=-\frac{GMm}{2a}
    \label{eq_vv}
\end{equation}
where $v$ is the orbital speed, $r$ is the instantaneous radial distance from the central body. Combined with \eqref{eq_v1'}, the radial component velocity could also be derived. An additional term for a high $e$ situation of $\vec{v_r}$ is then $\sqrt{v^2-|\vec{v_1}'|^2}  \hat{r}$ where $\hat{r}$ is the radial directional vector.

\subsubsection{Coverage model}

In the special case where both orbits are circular, each orbital alignment corresponds to a discrete opportunity for the target to enter the field of view of the optoelectronic sensors. This opportunity is determined by the relative velocity components, as well as by $F$ and $\Delta r$. The problem can therefore be regarded as analogous to a collision problem between the detection region and $U$. This formulation is also compatible with the framework proposed in \cite{Alfano2006}, which may be important in a more detailed detection-probability analysis. But we use the linear relative motion assumption in this model.

We first consider a scenario where the effects of the radial component of relative velocity are ignored, and provide a temporary conclusion. We will then theoretically derive how a target collides with our region of detection to consider the effects of the radial component of the relative velocity. The motion of the target is considered to be accurate in this theoretical calculation. Starting from $\alpha=\frac{\pi}{2}$, the perpendicular component of the relative velocity $u_1$ can be written as
\begin{equation}
    u_1=|\vec{v_1'}-\frac{\vec{v_1'}\vec{v_2'}}{|\vec{v_1}||\vec{v_2}|}\vec{v_2}|
    \label{eq_u1}
\end{equation}
and the parallel component of the relative velocity $u_2$ is 
\begin{equation}
    u_2=|\vec{v_2}-\frac{\vec{v_1'}\vec{v_2'}}{|\vec{v_1}||\vec{v_2}|}\vec{v_2}| .
    \label{eq_u2}
\end{equation}
The possibility of detection during a radial alignment event $P$ determined by the duty cycle when FoV$=F$ is:
\begin{equation}
P=\frac{\tan\frac{F}{2}\Delta r}{d}(\frac{u_2}{u_1}+1)
\label{eq_P}
\end{equation}
where $d$ is $2\pi(\rho+h_1)p/t$, the distance between adjacent satellites in the same orbital plane. 

Then we can extend this model to a more general case, i.e., the collision between the pyramid and $U$. Because $P$ is the central quantity in the integration and may vary with $\alpha$, we have to discuss how $v_r$and $\alpha$ will affect $P$. Since $P$ is determined by the duty cycle, the problem can then be reformulated as follows: for a given set of lines and a given square pyramid, how do the intersection constraints determine the feasible region of the lines?  Set the y-axis to be aligned with the direction of satellite motion, z-axis to be aligned with the radial direction, then:

\begin{equation}
    P=\frac{(y_{max}-y_{min})}{d}
    \label{eq_P'}
\end{equation}

Assume the directional vector within the pyramid is $\vec{j}=(j_x,j_y,j_z)$, the directional vector of the line of relative motion is $\vec{e}=(e_x,e_y,e_z)$ and $\Delta r$, and the corresponding y-axis coordinate is $\Delta r'$. Given the parameters that indicate two lines determined by the two directional vectors are $\tau, \tau'$ then:

\begin{equation}
    \tau\vec{j}=\tau'\vec{e}+\Delta r \hat{z}+\Delta r' \hat{y},
    \label{eq_je}
\end{equation}

by solving that we get:
\begin{equation}
    y=\Delta r'=\Delta r \frac{e_xj_y-j_xe_y}{e_xj_z-e_zj_x}
    \label{eq_y}
\end{equation}
This can be formulated as a linear programming problem in $\vec{j}$, whose maximum is attained on the boundary of the feasible region, specifically at the vertices. In the present case, these vertices are the four corner points of the intersection region between plane A determined by both $\vec{e}$ and the y-axis, and the square pyramid. After computing the length along the edges $l_1,l_2$, the normalized directional vectors are:

\begin{equation}
    \vec{j_{mn}}=A\frac{(\cos(\alpha+n\frac{F}{2}),\sin(m\frac{F}{2}),\sin(\alpha+n\frac{F}{2}))}{\cos(\frac{F}{2})}
    \label{eq_j}
\end{equation}
where $m,n$ are either $+1$ or $-1$, $A$ is the normalization constant. From \eqref{eq_y}, there is a relation$y_{nm}=y(j_{nm})$ for different pairs of $m,n$ where $y_{max}$ and $y_{min}$ could be obtained from $y_{nm}$. Therefore, we could use \eqref{eq_P'} to evaluate the possibility of detecting $U$ in an alignment. 

As discussed in \cite{s21113684}, Poisson-type models are commonly used in SSA to represent uncertain target populations and detection events within a sensor FoV. This therefore, provides a reasonable approximation for the present scenario. After determining $P$, the corresponding expected instantaneous rate of detection $\lambda_{inst}$ per unit time at this situation could be determined from:
\begin{equation}
    \lambda_{inst}=\frac{2P}{T_{0U}}
    \label{eq_lamb}
\end{equation}
where that coefficient of $2$ is from the two alignments per orbit cycle.

Since $U$ is chosen arbitrarily, its orbital plane is characterized by a right ascension of the ascending node (RAAN) $\phi$. Variations in $\phi$ change the intersection between the target orbit plane and the equatorial plane, and therefore also modify the intersection lines between the target orbit plane and the satellite orbit planes. 
For a group of these targets, the average revisit time is determined by:

\begin{equation}
T=\frac{\int  d\phi}{\int \frac{1}{T(\phi)}d\phi}
\label{eq_T}
\end{equation}

The number of observed targets over a short period oscillates around the mean value. The scanning period $T_3$ is determined by the time interval between equivalent instantaneous satellite spatial distributions in the same constellation configuration :
\begin{equation}
    T_3=\frac{T_0 p}{2t}
    \label{eq_T3}
\end{equation}
where $T_0$ is from \eqref{eq_T0}. By evaluating the equations \eqref{eq_T2} and \eqref{eq_T3}, it is always true that $T_3<T_2$. Therefore, $T_r$ given by \eqref{eq_Tr}is determined by the Poisson expected revisit period $T$, and the beat period $T_2$.

The stationary coverage, defined as the coverage at any given instant, is the ratio of the detection volume to the volume of the target space.
\begin{equation}
    C_s=\frac{\frac{4}{3}tw^3 \tan^2{\frac{F}{2}}}{\frac{4}{3}\pi(r_2^3-r_1^3)}=\frac{tw^3\tan^2{\frac{F}{2}
    }}{\pi(r_2^3-r_1^3)}
    \label{eq_Css}
\end{equation}

To evaluate an SSA system of this kind, three metrics are considered:  $T_r$ from \eqref{eq_Tr},  $C_a$ from \eqref{eq_Ca}, and $C_s$.

\subsubsection{Air Drag}

Orbital decay for a satellite depends on its ballistic coefficient $B$. For a symmetric constellation, the geometric configuration therefore evolves uniformly. In the present analysis, all satellites are assumed to follow circular orbits. According to the virial theorem, the kinetic energy, $K.E.$, is defined as:
\begin{equation}
    K.E.=-\frac{1}{2}\Sigma(\vec{F}_k \vec{r}_k)
    \label{eq_KE}
\end{equation} 
where $F_k$ is the $k$th force, $\vec{r_k}$ is the $k$th position vector.
For a gravitationally bound orbit, $2K.E.=-V$ where $V$ is the potential energy. From the conclusions in \cite{goldstein2002}, the energy of the system $E$ is:
\begin{equation}
    E=-\frac{GMm}{2a}
    \label{eq_Eg}
\end{equation}
where $a$ is the semi-major axis equals to $\rho+h_0$. The rate of descent in energy is the product of air drag and velocity. Define $\vec{f(v)}$ as the air drag per unit mass at velocity $v$, then a equation could be determined as:
\begin{equation}
    \frac{GMm}{2a^2} \frac{da}{dt}=-\vec{f}(\sqrt{\frac{GM}{a}})\sqrt{\frac{GM}{a}}m.
    \label{eq_gm}
\end{equation}

The air drag per unit mass is 
\begin{equation}
    \vec{f}=-\frac{\rho_{air}(h)v_{ra} \vec{v_{ra}}}{2B}
    \label{eq_fr}
\end{equation}
where $B=\frac{m}{C_D A}$, $C_D$ is the coefficient of air drag, $A$ is the area, $m$ is the mass of a satellite. $B$ of every satellite are the same, the only two variables are $v_{ra}$ and $\rho_{air}(h)$. $v_r$ is the difference between the velocity of the satellite and the velocity of air from Earth's rotation. $\rho_{air}(h)$ could be derived and approximated from high precision air model such as the NRLMSISE-00 model \cite{Picone2002}. Although the atmospheric mass density at a given altitude may be influenced by many factors, the orbital period in LEO is sufficiently short that interpolation and a periodic approximation can be used in this model.

Using the gravitational parameter $\mu=\sqrt{GM}$ and summarizing \eqref{eq_gm}--\eqref{eq_fr}, we get the following:
\begin{equation}
    \frac{da}{dt}=-\frac{\sqrt{\mu}}{B}\rho_{air}(a)a^{\frac{1}{2}}
\label{eq_da}
\end{equation}

Due to symmetry, satellites within the same constellation are expected to undergo similar orbital decay, thereby forming a decaying shell. Air drag does not induce any precessional angular velocity \cite{strizzi1993improved}. Under a local approximation, the orbital decay may then be represented as the solution to \eqref{eq_da}, which is a polynomial function of time.

\section{Proposed Algorithm}
For an arbitrary target $U$ with semi-major axis $a$, inclination $i$ and linear eccentricity $e$, consider the family of elliptical orbits associated with these parameters. For such a family of ellipses, the velocities at each given point define a vector field that depends only on the radius. According to conservation of angular momentum, the transverse velocity could be approached from \eqref{eq_v1'}. Instead of directly integrating different elliptical orbits, we may treat $U$ equivalently as a family of imaginary circular orbits for the calculation mathematically, although these auxiliary circular orbits do not themselves satisfy the orbital dynamics of the original system.

According to orbital dynamics, we have
\begin{equation}
    r=\frac{a^2-c^2}{a+c \cos{\theta}}
    \label{eq_r}
\end{equation}
where $c=ea$ is the semi-focal length
If the detection process is approximated as a Poisson process, then over an infinitesimal time interval $dt$, corresponding to an angular displacement $d\phi=\frac{v_t}{r}dt$ the mean detection probability for the constellation is obtained by summing the probabilities associated with each interception of the constellation orbits over the corresponding cycle $T(\phi)=\frac{2\pi r}{v_t}$. For each time $U$ and the satellites align radially, $p$ could be approached from $eq_P$ with different parameters. Since the alignment may occur at any phase along the orbit, the expectation is obtained by averaging over phase, which is a region of a circle in the velocity field. In addition, the relative right ascension of the ascending node (RAAN) phase between the constellation and the target may vary, which requires an additional averaging step. Let that variable be $\varphi_{or}$. The equation is:
\begin{equation}
\lambda=2\times\frac{\int Average(\frac{\Sigma(P)}{T(\phi)})\frac{r}{v_t}d\phi}{\int dt}
\label{eq_la}
\end{equation}

After simplification, this expression reduces to an average over the coordinate $\phi$. This may be regarded as the elliptic-orbit variation of \eqref{eq_lamb}.

To facilitate our subsequent discussion, we list the definitions of variables used in the proposed Algorithm \ref{alg1} in Table~\ref{tab:var_definitions}.

\begin{table}[H]
    \centering
    \caption{Algorithm Variables}
\resizebox{0.86\columnwidth}{!}{%
\begin{tabular}{ll}
\toprule
$\xi$ & Angular variable for integration \\
$K$ & step number that controls all the steps in the algorithm\\
$r$ & Distance from the center of the Earth of the elliptical orbit\\ 
$\rho_0$ & The Earth radius\\
$\rho$ & The constellation radius\\
$\varphi_{or}$ & Relative RAAN between the target and constellation.\\
$k$ & Numpy array for $p$ planes\\
$p_{ro}$ & Accumulator for multiple orbital planes\\
$p_{bo}$ & Accumulator for different $\varphi_{or}$ \\
$p_{to}$ & Accumulator for different angle\\
$P$ & Possibility from \eqref{eq_P}\\
$\Delta_k$ & Plane offsets\\
$\Theta_{i,k}$ & Angle grid for different $\varphi_{or}$ and $\Delta_k$\\
$P_{i,k}$ & Probability for different $\varphi_{or}$ and $\Delta_k$ from \eqref{eq_P}\\
$c_{i,k}$ & Inner product of $\vec{v_1'}$ and $\vec{v_2'}$ for different $\varphi_{or}$ and $\Delta_k$\\
$c^2_{i,k}$ & $c_{i,k}\times c_{i,k}$\\
\bottomrule
\end{tabular}
}
\label{tab:var_definitions}
\end{table}
\subsection{Algorithm summary}

\begin{algorithm}[htbp]
\caption{Poisson expected revisit period}
\begin{algorithmic}[1]

\STATE \textbf{Initialize} set the step number variable $K$.
\STATE \textbf{Initialize} phase samples $\{\varphi_{or}^{(i)}\}_{i=1}^{K}$ and plane indices $k=0,\dots,p-1$.
\STATE \textbf{Precompute} plane offsets $\Delta_k := k\cdot \frac{2\pi}{p}$.
\STATE \textbf{Precompute} the 2D angle grid $\Theta_{i,k} := \varphi_{or}^{(i)} + \Delta_k$.
\STATE \textbf{Precompute} geometry-only terms $c_{i,k}$ (and optionally $c_{i,k}^2$) using \eqref{eq_u1} and \eqref{eq_u2}.
\STATE \textbf{Initialize} accumulator $p_{to}=0$
\STATE \textbf{Initialize} multiply factor $\gamma_0$ and $\delta_0$ equal to $\sin{F}$ in \eqref{eq_P}
\WHILE{$r-\rho \le w$ \AND $\xi < \pi$}
    \STATE \textbf{Advance} $\xi$ by a fixed step.
    \vspace{0.3em}
    \STATE \textbf{Given} current geometry $r=r(\xi)$ and $\Delta r=r-\rho$.
    \IF{$\Delta r < 0$}
        \STATE \textbf{Return} $p_{bo}=0$.
    \ELSE
        \STATE \textbf{Evaluate} dimensionless velocities $m_v,n_v$ using \eqref{eq_mv} and \eqref{eq_nv}.
        \STATE \textbf{Evaluate} range-scaled parameters $\gamma := \gamma_0\Delta r$, $\delta := \delta_0\Delta r$.
        \STATE \textbf{Evaluate} geometric quantities on the $(i,k)$ grid:
        \STATE \hspace{1.5em} $v_{1,i,k} \leftarrow v_1(m_v,n_v,c_{i,k})$ using \eqref{eq_v1'}
        \STATE \hspace{1.5em} $v_{2,i,k} \leftarrow v_2(m_v,n_v,c_{i,k})$ using \eqref{eq_v2'}
        \STATE \hspace{1.5em} Find the third component $v_{3,i,k}$ using \eqref{eq_vv}

        \STATE \textbf{Evaluate} a vectorized condition $\mathbb{M}_{i,k}$ whether $P>=1$.
        \STATE \textbf{Evaluate} contribution $P_{i,k}$ using \eqref{eq_P} .

        \STATE \textbf{Sum} over both sampling directions:
        \STATE \hspace{1.5em} $p_{ro}:= \sum_{i=1}^{K}\sum_{k=0}^{p-1}P_{i,k}$.
        \STATE \textbf{Normalize} and output:
        \STATE \hspace{1.5em} $p_{bo} := \dfrac{p_{ro}}{K\cdot 2\pi}$.
        \STATE \textbf{Return} $p_{bo}$.    
    \ENDIF
    \STATE \textbf{update} accumulator $p_{to}$ with normalization coefficients $\frac{\pi}{K}$ or $d\xi$
    \STATE \textbf{Update} the geometry-dependent quantity $r$ \eqref{eq_r}.
\ENDWHILE
\STATE \textbf{Evaluate} the characteristic time scale using \eqref{eq_T}, \eqref{eq_la} and conclusions in Poisson process.
\STATE \textbf{Compute} the final rate and the corresponding time.
\STATE \textbf{Evaluate} the additional derived time scale from \eqref{eq_T1}.
\end{algorithmic}
\label{alg1}
\end{algorithm}

Our proposed $T$ can be approached from Algorithm \ref{alg1}, which gives a function $T(U,S)$, where $S$ represents the geometric quantities of the constellation. The definitions of parameters is in table \ref {tab:var_definitions}.  With \eqref{eq_T2}, we can determine $T_r$ from:
\begin{equation}
    T_r=T_r(U,S)=Max(T(U,S),T_2)
    \label{eq_Tr'}
\end{equation}
\ref{alg1}, if there are multiple types of constellations:
\begin{equation}
    T_{r-sum}=SUM({T_r(U,S)})_S
    \label{eq_Trs}
\end{equation}
for all the objects within a region:
\begin{equation}
    T_{r-ave}=Ave({T_r(S)})_U
    \label{eq_Tra}
\end{equation}
Which is an average over all possible objects, and also
\begin{equation}
    \frac{dT_r}{dt}=\frac{dT_r(U,S)}{dS}\frac{dS}{dt}
    \label{eq_Trd}
\end{equation}

where the orbital decay rate $\frac{da}{dt}$ from \eqref{eq_da} is an implementation of \eqref{eq_Trd}. The function $T_r(U,S)$ is a key descriptive quantity, from which the desired metrics can be obtained through differentiation or integration.

\subsection{Algorithm evaluations}

The computational cost of this algorithm is primarily governed by two parameters: the number of discretization steps, denoted by $K$, and the number of orbital planes, denoted by $p$. According to the structure of the algorithm, the evaluation of \eqref{eq_Tr'} has time complexity $O(K^2)$. The summation over satellites indexed by $i:t/p/f$ results in a time complexity $O(p)$. Each additional calculus operation applied to \eqref{eq_Tr'} increases the order of the complexity with respect to $K$. 

The algorithm involves two numerical integrals that both depend on the chosen step size. Since the integration domains are different, the optimal number of steps for achieving the highest accuracy is generally not the same for the two integrals. For a common step size, that theoretical error is proportional to 
\begin{equation}
    Error=\frac{D}{K}+D_1+D_2 K+D_3K^2
    \label{eq_err}
\end{equation}
where $D_i$ are determined by other parameters and the accuracy of the datatype. In practice, $K=500$ is sufficient to ensure that the numerical error remains acceptable.

\section{Numerical Results}
\hl{This section compares the proposed model with existing models and discusses their advantages and limitations. Through simulation, the influence of geometric parameters on the performance of Walker--Delta constellations in SSA missions are clarified. Case studies are presented for both existing constellations and a hypothetical constellation. In addition, orbital decay is used as an example to demonstrate how the proposed algorithm can evaluate the effects of additional influencing factors.}

The evaluation was conducted on the high-performing computing (HPC) facility where the Intel 6972P CPU (with 2.4 GHz 384 MB cache L3) with single core configured 4 GB RAM are used in the experimentation.

\subsection{Comparison with the existing model}
The models proposed in \cite{2Anniballe2025} and \cite{xie2022} are selected for comparison. Where necessary, adaptations are made to account for unspecified parameters in prior studies and for parameters that are not universally applicable across all models, ensuring a consistent basis for comparison.

\begin{table}[htp]
\centering
\caption{Representative model comparison for debris observation in LEO.}
\label{tab_model-comparison}
\setlength{\tabcolsep}{4pt}
\renewcommand{\arraystretch}{1.15}
\begin{tabular}{@{}p{0.22\linewidth} p{0.74\linewidth}@{}}
\toprule
\textbf{Model} & \textbf{Main assumptions, modeling strategy, and outputs} \\
\midrule

Xie \textit{et al.} (2022)\cite{xie2022}
&
\textbf{Time model:} deterministic revisit model based on repeated strip-scan coverage. \newline
\textbf{Geometry:} scan-band coverage; uses daily laps $N$ and scan width $D=2\,\mathrm{AOFOV}$. \newline
\textbf{Core derivation:} represent $N=K+m/M$ and define revisit period $M$; require $(KM+m)D\ge 360^\circ$ and obtain $M_{\min}$. \newline
\textbf{Primary output:} $M_{\min}$ (days), i.e., deterministic revisit period for a differential orbital element. \newline
\\
\midrule

D'Anniballe \textit{et al.} (2025)\cite{2Anniballe2025}
&
\textbf{Time model:} simulation-based statistical anonymous revisit analysis.  \newline
\textbf{Visibility:} orbital visibility is determined by multiple physical and observational constraints, including Earth shadowing, horizon/background limits, field of regard, field of view, and limiting magnitude. \newline
\textbf{Primary outputs:} coverage metrics + expected revisit time / revisit-time distributions for uncatalogued millimetre-sized debris. \newline
\\
\midrule

Proposed model
&
\textbf{Time model:} stochastic revisit via Poisson process with mean detection rate $\lambda$ and expected revisit time $T_r$.  \newline
\textbf{Geometry:} explicit FoV constraint combined with work-distance and relative-orbit geometry \newline
\textbf{Primary outputs:} $\lambda$, $T_r=\mathbb{E}[\tau]=1/\lambda$ (days) for recognizable targets. \newline
\textbf{Computation:} Analytical evaluation without time-step physical simulation.
\\

\bottomrule
\end{tabular}
\end{table}

Table \ref{tab_model-comparison} illustrates the main differences between different models. In this work, we employ an analytic approach to estimate the time-averaged performance metrics of a given constellation. The simulation-based methods \cite{2Anniballe2025} capture realistic visibility constraints, but rely on simulation and are less suited for fast analytical parameter sweeps. The proposed model is analytic and is expected to approach convergent results with low computational cost. Another important difference between the proposed model and \cite{2Anniballe2025} is whether the targets are recognizable, which is vital for SST tasks in SSA.
The formulation of the proposed model is also readily adaptable---by modifying the underlying assumptions to match specific scenarios, the same framework can be extended to examine more complex relationships between system parameters and performance metrics. Xie's model \cite {xie2022}, however, focuses on a specific scenario and implements less universal assumptions.

\paragraph{Further comparison to Xie's model:}

Xie's model \cite{xie2022} let $T_g$ denote the length of one day and $T$ the orbital period of the observing satellite. The number of satellite revolutions per day is
\begin{equation}
N=\frac{T_g}{T}.
\label{eq_xieN}
\end{equation}
Decompose $N$ into its integer and fractional parts:
\begin{equation}
N = K + \frac{m}{M}, \qquad 0<m<M,\ \gcd(m,M)=1,
\label{eq_xiedecomp}
\end{equation}
where $K$ is the integer part and $m/M$ is the fractional part. Define the per-revolution scan (or coverage) width on a one-dimensional angular domain as
\begin{equation}
D = 2\,\mathrm{AOFOV}.
\label{eq_xieD}
\end{equation}
Over $M$ days, the total number of revolutions equals
\begin{equation}
MN = M\left(K+\frac{m}{M}\right)=KM+m.
\label{eq_xietl}
\end{equation}
Under the idealized assumption that successive scan strips have negligible overlap, the accumulated covered angular width over $M$ days is approximately $(KM+m)D$. Requiring full coverage of a $360^\circ$ angular domain yields the condition
\begin{equation}
(KM+m)D \ge 360^\circ.
\label{eq_xie_cover_condition}
\end{equation}
Solving \eqref{eq_xie_cover_condition} for $M$ gives
\begin{equation}
M \ge \frac{360^\circ}{K D} - \frac{m}{K},
\end{equation}
and therefore the minimum revisit (coverage) period in days is
\begin{equation}
M_{\min}=\left\lceil \frac{360^\circ}{K D}-\frac{m}{K}\right\rceil,
\label{eq_xie_Mmin}
\end{equation}
which can also be written (equivalently) as
\begin{equation}
M_{\min}=\left\lfloor \frac{360^\circ}{K D}-\frac{m}{K}\right\rfloor+1.
\label{eq_xie_Mmi}
\end{equation}

To adapt Xie's model and compare it with our model, we choose a specific case where the targets are not in GEO but in close LEO, where $K=1$. The adaptive Xie's model is:
\begin{equation}
    M_{min}^*=\frac{360^\circ}{\frac{2\Delta r\tan(\frac{F}{2})}{\rho+h_1}\times 2}
    \label{eq_mm}
\end{equation}

and the corresponding time is:
\begin{equation}
    T_{min}^*=M_{min}^*\times\frac{2\pi}{\sqrt{\frac{GM}{\rho}}}
    \label{eq_tm}
\end{equation}

This corresponds to the adaptive Xie's model. In the present implementation, we neglect the incomplete-orbit term in \eqref{eq_xie_Mmi} equal to $\left\lfloor M_{min}^* + \frac{m}{K} \right\rfloor+1-M_{min}^*$. This term represents the time before the next observation.  In addition, one geometric simplification in Xie's original formulation should be noted: the FOV is modeled as a zero-thickness triangle rather than a pyramidal volume, resulting in the absence of two bounding surfaces. For most orbits, the relative velocity generally has a component that is perpendicular to these two planes. This factor will affect the scanning rate as explained in \eqref{eq_P}, \eqref{eq_P'}, \eqref{eq_j}. A rough approximation of this factor for LEO close-range observation is an additional coefficient of $2$ according to \eqref{eq_P} in our model. The adjusted model is named Xie* model.

To compare the results of the models, both models are implemented to the scenario where $w = 100$ km, while varying $\Delta r$. The spacing of the parameter $\Delta r$ decreases near both endpoints of the interval. This choice covers the full feasible region under the condition of $w = 100$ km. This scenario focuses on the interaction between a single satellite and a single target $U$ for clearity.

\begin{table}[htbp]
\caption{Revisit time (days for 1 satellite) $w=100$ km} 
\begin{center}
\begin{tabular}{c | c c c } 
 \toprule
 $h_1-h_0$ & Xie adaptive & Xie* & Poisson (Ours) \\ 
 \midrule
 2000 & 29096 & 14548 & 14544\\ 
 \hline
 6000 & 9704.2 & 4852.1 & 4848.9\\
 \hline
 10000 & 5825.86 & 2912.93 & 2909.74\\
 \hline
 20000 & 2917.13 & 1458.56 & 1455.39\\
 \hline
 40000 & 1462.76 & 731.38 & 728.21\\
 \hline
 60000 & 977.97 & 488.99 & 485.82\\
 \hline
 80000 & 735.58 & 367.79 & 364.62\\
 \hline
 90000 & 654.78 & 327.39 & 324.22\\
 \bottomrule
\end{tabular}
\label{tab1}
\end{center}
\end{table}

Table \ref{tab1} presents the computed revisit time given by two adaptive models and the proposed model across different parameter sets. There is high consistency between the Xie* model and our Poisson model. There are still deviations because of the use of a fixed correction coefficient. This is showing a high precision of evaluation.

This result can be naturally extended to multi-satellite constellations. In Xie’s closed-form model, if the residual incomplete half-orbit before the next observation opportunity is neglected, the revisit time is inversely proportional to the total number of satellites. For a $0^\circ$ inclination target, the Walker-star constellation has geometrical symmetric orbital planes, so the revisit time is bound to be inversely proportional to the number of orbital planes $p$:
\begin{equation}
    T_r=\frac{T_{const}}{t}
    \label{eq_Trt}
\end{equation}

Under the sparse-constellation assumption where there is no overlapping, satellites within each plane can further be regarded as statistically independent and equivalent, so the revisit time is also inversely proportional to the number of satellites in an arbitrary orbital plane. Therefore, for Walker-delta constellations observing $0^\circ$ inclination targets, the revisit time is inversely proportional to the total satellite number $t$. Accordingly, Table \ref{tab1} is extendable by scaling the total number of satellites and inversely scaling the revisit time. 

The results indicate that our model and Xie’s model share a consistent underlying mechanism. In particular, both approaches capture the same geometric constraints determines the observation process, leading to similar scaling behavior of the revisit time. Under specific simplifying assumptions, the proposed model can be reduced to a closed-form expression, which is consistent with the formulation of Xie’s model. This further confirms that both approaches are determined by the same underlying mechanism, while our model provides a more general framework beyond these idealized conditions.
Xie’s model adopts a closed-form expression, which provides clear analytical insight but relies on simplified geometric assumptions. In contrast, our model which is still analytical, does not require a closed-form representation and can be extended to a broader class of scenarios, including general Walker-delta constellations, arbitrary target orbits, non-stationary configurations, and detection regions with complex geometries.
Therefore, while Xie’s model offers a closed form solution under idealized conditions, the proposed model provides greater flexibility for handling more general and realistic observation scenarios, while remaining an analytical description.

To assess the computational performance of the proposed model, the execution time was also measured. The runtime behavior depends on the coding formulation in addition to the theoretical algorithmic complexity. The case chosen is:

$i=\frac{\pi}{2}:t=36\times p/p=2^n, n\in[0,7]/f$

$\Delta r=2000$ m

$w=3000$ m

Target orbit inclination:$0^\circ$ 

We choose exponentially increasing values of $P$, which is the code variable represents $p$. This choice is motivated by the dominance of lower-order terms at small $P$. $K$ is chosen uniformly within $[400,3200]$ with an additional $K=200$ to illustrate the full transition by which the dominant term $K^2 p$ becomes the majority as $K$ and $p$ increase.

\begin{table*}[htbp]
\caption{Relationship between time cost per second and step number $K$, plane number $P$.}
\label{tab3}
\centering
\resizebox{1.8\columnwidth}{!}{
\small
\begin{tabular}{c|c|c|c|c|c|c|c|c}
\toprule
 {$K$ vs $P$ } & {$P$=1} & {$P$=2} & {$P$=4} & {$P$=8} & {$P$=16} & {$P$=32} & {$P$=64} & {$P$=128} \\
 \midrule
 200  & $3.6 \pm 0.5$      & $3.357 \pm 0.012$ & $3.821 \pm 0.006$ & $4.69 \pm 0.13$ & $6.221 \pm 0.008$ & $9.58 \pm 0.04$ & $18.5 \pm 2.2$ & $54.6 \pm 2.9$ \\
 \hline
 400  & $6.439 \pm 0.027$  & $7.409 \pm 0.022$ & $8.94 \pm 0.04$ & $12.3 \pm 0.6$ & $18.58 \pm 0.15$ & $34.3 \pm 1.4$ & $108 \pm 11$ & $216 \pm 13$ \\
 \hline
 800  & $14.797 \pm 0.022$ & $20 \pm 4$ & $24.177 \pm 0.019$ & $37.4 \pm 0.5$ & $70.5 \pm 2.9$ & $204 \pm 4$ & $422 \pm 14$ & $834 \pm 9$ \\
 \hline
 1200 & $24.50 \pm 0.21$   & $31.8 \pm 0.5$ & $46.4 \pm 0.6$ & $78.8 \pm 2.3$ & $157 \pm 13$ & $294 \pm 13$ & $969 \pm 6$ & $2011 \pm 9$ \\
 \hline
 1600 & $36.0 \pm 0.7$     & $48.5 \pm 0.5$ & $75.0 \pm 0.7$ & $140 \pm 8$ & $272 \pm 22$ & $514 \pm 8$ & $1661 \pm 10$ & $3680 \pm 30$ \\
 \hline
 2000 & $48.9 \pm 0.7$     & $69.0 \pm 0.7$ & $111.7 \pm 0.9$ & $220 \pm 18$ & $421 \pm 23$ & $823 \pm 14$ & $2712 \pm 20$ & $5695 \pm 15$ \\
 \hline
 2400 & $63.4 \pm 0.7$     & $91.8 \pm 0.5$ & $163 \pm 8$ & $309 \pm 14$ & $573 \pm 10$ & $1196 \pm 13$ & $3980 \pm 14$ & $8158 \pm 30$ \\
 \hline
 2800 & $79.5 \pm 1.9$     & $119.1 \pm 1.9$ & $217 \pm 12$ & $409 \pm 27$ & $775 \pm 4$ & $1634 \pm 5$ & $5410 \pm 26$ & $12172 \pm 27$ \\
 \hline
 3200 & $97.0 \pm 0.7$     & $148.7 \pm 0.9$ & $277 \pm 11$ & $542 \pm 30$ & $1018 \pm 4$ & $2188 \pm 7$ & $7248 \pm 40$ & $14839 \pm 26$ \\
 \bottomrule
\multicolumn{9}{l}{in ms, reported as mean $\pm$ standard deviation. Each experiment is repeated 5 times with 1 warm-up run excluded.}
\end{tabular}
}
\end{table*}

Table \ref{tab3} shows timing performance in ms, which is consistent with our complexity analysis. For small to moderate parameter settings, the runtime is still influenced by implementation-level lower-order effects associated with the NumPy-based formulation. The dominant theoretical term becomes much more apparent at Large $K$ and $p$. Overall, the results indicate that the dominant theoretical term $K^2 p$ becomes progressively more visible in the large-$K$, large-$p$ regime, consistent with the theoretical complexity. Furthermore, the runtime stays within a practically manageable level, indicating that the model is suitable for repeated evaluation and parameter exploration.

\subsection{Geometric parameters}
In this section, the analysis focuses on how the total number of satellites, the number of planes, and the inclination influence the revisit time as well as the detection gap, and verifies if the results are consistent with theoretical results \eqref{eq_Trt}.

In a case where:
Step number $K$=500

Step size for n:1. $p=2^n$,$t=36p$.

$i=\frac{\pi}{2}:t=36\times p/p=3,6,9/f$

$\Delta r=2000$m

$w=3000$m

Target orbit inclination:$0$ 

For a single-layer constellation, the revisit time T in days can be computed with different values of $t$ and $p$. We choose uniformly distributed $t\in[36,180]$ to output moderate results and avoid any unnecessary errors. Uniformly distributed $p \in [3,9]$ is also complete to show the regular pattern.
\begin{table}[htbp]
\caption{Revisit time at different $t$ and $p$ values}
\begin{center}
\begin{tabular}{c | c c c} 
 \toprule
 t & $p$=3 & $p$=6 & $p$=9 \\ 
 \midrule
 36 & 12.12 & 12.12 & 12.12 \\ 
 \hline
 72 & 6.060 & 6.060 & 6.060 \\
 \hline
 108 &3.030  & 3.030 & 3.030 \\
 \hline
 144 & 2.020 & 2.020 & 2.020 \\
 \hline
 180 & 1.515 & 1.515 & 1.515 \\ 
 \bottomrule
\end{tabular}
\label{tab1.1}
\end{center}
\end{table}

In Table \ref{tab1.1}, the revisit time calculated is inversely proportional to the total number of satellites. This Table \ref{tab1.1} also suggests that the number of orbital planes $p$ influences the revisit time $T_r$ by a neglectable amount. That is consistent with our analysis in the model comparison\eqref{eq_Trt}.

Another important parameter that might affect the performance is $i$. Take several different $i$ with step=$15^{\circ}$ .
\begin{table}[htbp]
\caption{Relationship between revisit period and inclination}
\begin{center}
\begin{tabular}{c | c} 
 \toprule
 \textbf{Constellation Inclination (degree)} & \textbf{$T$ (days)}\\
 \midrule
 15 & 21.39 \\ 
 \hline
 30 & 19.11\\
 \hline
 45 & 17.14 \\
 \hline
 60 & 15.37 \\
 \hline
 75 & 13.72 \\
 \hline
 90 & 12.12 \\
 \bottomrule
\end{tabular}
\label{tab2}
\end{center}
\end{table}
Table \ref{tab2} presents the average revisit time of target $U$ as a function of target inclination. The results show that $T$ decreases with increasing inclination.

In this subsection, $T$ is shown to depend primarily on $t$ and $i$. According to \eqref{eq_T2} and \eqref{eq_T3}, both $T_2$ and  $T_3$ decrease as $p$ is decreases, as they are proportional to $p$. In addition, $T$ is reduced at higher target inclinations. Therefore, within the present framework, smaller $p$ and larger $i$ generally correspond to shorter $T_r$ according to \eqref{eq_Tr}.

\section{Case Study}
\subsection{Implementations on current constellations}

In this subsection, we examine the performance of a Starlink-like constellation for two specific yet representative targets $U_1$ and $U_2$. Both are inclined to $i_U=0$ .

For clearity, We use a Starlink Group 1 \cite{Gunter_page} shell at 540–550 km with $i\approx53.0^{\circ}$, consisting of 72 orbital planes with 22 satellites per plane, for a total of 1584 satellites \cite{Gunter_page,2026Star}. The key parameters are listed in Table \ref{constellation_params} and are consistent with SpaceX's FCC-authorized configuration and public sources \cite{2026Star}. A better explanation is Starlink (Gen1) operational shell at $540–550$ km, $i\approx 53.0 ^\circ$, 72 planes, 22 sats/plane (1584 sats).

\begin{table}[htbp]
\caption{Constellation Parameters for Experimentation}
\begin{center}
\resizebox{0.88\columnwidth}{!}{%
\begin{tabular}{l|c|c|c|c}
\toprule
 {Constellation} & {$i$} & {$t$} & {$p$}&{$h_0$}\\
 \midrule
 {Starlink} & 53.0 & 1584&72  & 550 (km)\\ 
 \hline
  {Made up constellation} & 90 & 288 & 8  & 550 (km)\\ 
 \bottomrule
\multicolumn{4}{l}{}
\end{tabular}
}
\label{constellation_params}
\end{center}
\end{table}

Table \ref{constellation_params} also presents an additional hypothetical constellation that will be discussed in this section. As shown in the previous analysis, the inclination of a Walker-delta constellation plays an important role in determining $T$. Moreover, for a fixed total number of satellites, increasing $p$ not only raises the computational cost but also leads to a less homogeneous detection pattern for a fixed total number of satellites.

\begin{table}[htbp]
\caption{Three representative cases for Starlink group1 and one case for a Starlink group1 variation}
\centering
\resizebox{0.88\columnwidth}{!}{%
\begin{tabular}{c|c|c|c|c|c}
\toprule
 {Case No.} & {$p$} & {AA-$h_0$}&{AP-$h_0$}&{$T$}&{$T_2$}\\
 \midrule
 1 & $72$&$2$ km & $2$ km & $0.3675$ days & $6.98$ days\\ 
 \hline
 2 & $72$&$9$ km& $2$ km &$1.507$ days & $2.54$ days \\
 
 \hline
 3 & $72$&$100$ km& $2$ km &$5.704$ days & $0.276$ days \\
 
 \hline
 4 & $22$&$2$ km& $2$ km &$0.3675$ days & $2.13$ days\\
 
 \bottomrule
\multicolumn{4}{l}{}
\end{tabular}
\label{tab4}
}
\end{table}

As shown in Table \ref{tab4}, the Starlink constellation exhibits limited performance in this case. The main reason is that $T_2$ remains larger than $T$, and $T_r$ is estimated by \eqref{eq_Tr}. The total number of satellites is no longer the dominant factor. Instead, because the constellation is distributed over too many orbital planes, the number of satellites per plane becomes insufficient to ensure stable observation opportunities at successive orbital alignments. When the frequencies are close, a long beat period can arise due to interference between the corresponding components. Of course, in the low-$p$ regime, $P$ in \eqref{eq_P'} may also reach $1$. In that case, part of the satellite capacity becomes redundant for decreasing the revisit time.

\subsection{A custom constellation}
Here we define a custom constellation whose parameters are listed in Table \ref{tab4}.

From previous discussions, a fully optimized constellation consists of multiple polar orbital planes $i=\frac{\pi}{2}=90^{\circ}$. To avoid collisions and be representative, this hypothetical constellation is designed as in (\ref{constellation_params}).

According to the results in Table \ref{tab1.1} and the relation in \eqref{eq_Trt}, a total satellite number of 288 can reduce the computational cost while still providing an appropriate revisit time, with $T_2$ remaining smaller than $T$. $AA-h_0$ is set in three values of $2,9,100$ km, each represents a type of orbit. $AA-h_0=2$ represents a circular orbit. $AA-h_0=9$ is very close to a circular orbit, but can be outside the work region. $AA-h_0=100$ represents an orbit that intersects the detection region less frequently. 

\begin{table}[htbp]
\caption{Three representative cases}
\begin{center}
\resizebox{0.88\columnwidth}{!}{
\begin{tabular}{c|c|c|c|c}
\toprule
 {Case Number} & {AA-$h_0$}&{AP-$h_0$}&{$T$}&{$T_2$}\\
 \midrule
 1 & $2.00$ (km) & $2.00$ (km) & $1.516$ (days) &$4.263$ (days)\\ 
 \hline
 2 & $9.00$ (km)& $2.00$ (km) &$6.219$ (days)&$1.551$ (days)\\
 \hline
 3 & $100.0$ (km)& $2.00$ (km)&$23.22$ (days) &$0.169$ (days)\\
 \bottomrule
\multicolumn{4}{l}{}
\end{tabular}
\label{tab5}
}
\end{center}
\end{table}

From Tables \ref{tab4} and \ref{tab5}, it can be found that for a low work distance constellation, the revisit time changes rapidly as AA increases, provided that the fixed PA remains within the effective working range of the optical sensors. This is not always true and could change if $w$ is changed or at a even smaller PA.

\subsection{Work distance $w$ influences}

At high work distance $w>100$ km, some of the approximations might not be accurate enough. Interestingly, a typical extreme of optoelectronic sensors is around $w=100$ km. In previous work \cite{Michel2023}, which shows an example of $200$ mm aperture diameter on a 4700 $\times$ 3500 px sensor, suggesting a $100$ mm aperture diameter in our case. In the classical result \cite{Rayleigh1879} $\theta=1.22\frac{\lambda}{D}$, the physical resolution limit at $100$ km is close to $0.7$ m. 

In this section we use $s$ to represent the data of $s=AA-h_0$, where $AA$ is the apogee altitude and all units in kilometers (km).

At this stage, the coverage of availability must be adjusted. We have \eqref{eq_Ca} that:
\[
    C_a=1-\frac{\Sigma (\rho_{i+1}-\rho_{i})^2}{(r_2-r_1)^2}
\]

Now, a restriction $r_i>h_0+w$ or $r_{i+1}<h_0$ must be added when performing the sum.

The physical limit is restricted by the image sensor due to the finite number of pixels. This type of restriction restricts the linear resolution, which means that the length $l'=w\sin{\frac{F}{2}}$ is conserved. 
\begin{table}[htbp]
\caption{$T$ vs. Apogee Altitude $s$ and work distance $w$ at Perimeter Altitude $r_1=h_1=2000$ m}
\begin{center}
\begin{tabular}{c|c|c|c} 
 \toprule
 {$s$/km } & {$w=3000$ (m)} & {$w=8000$ (m)}&{$w=50000$ (m)}\\
 \midrule
 2 & 1.516 & 4.042 & 25.26\\ 
 \hline
 9 & 6.219 & 2.475 &9.201\\
 \hline
 100 & 23.48 & 12.87 &5.324\\
 \hline
 1000 & 76.14 & 43.46 &20.59\\
 \bottomrule
\end{tabular}
\label{tab6}
\end{center}
\end{table}

\begin{figure}[H] 
	\centering 
	\includegraphics[width=0.5\textwidth]{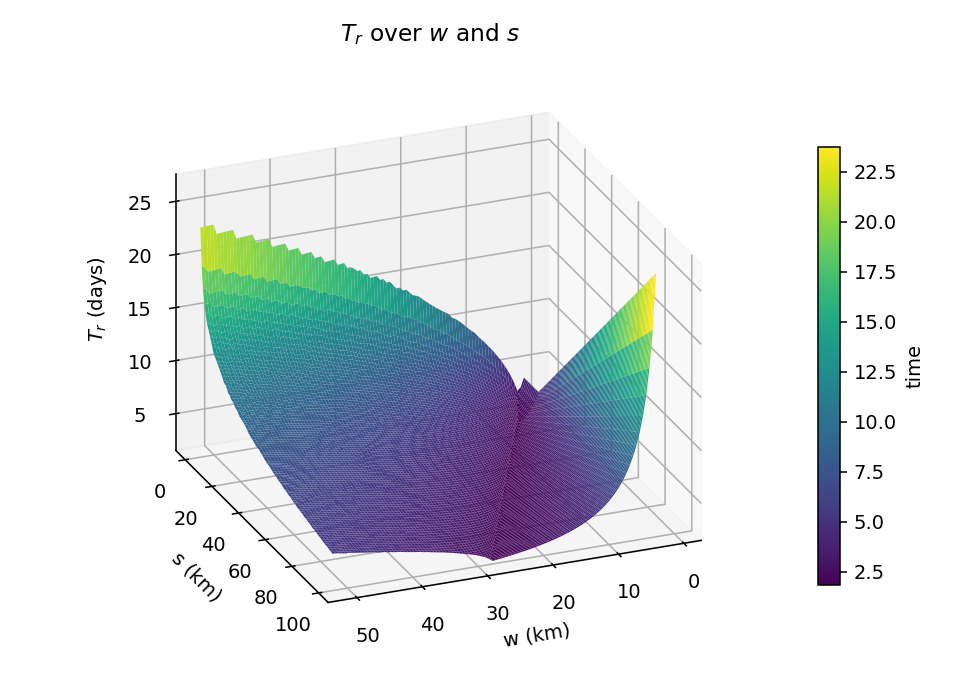} 
	\caption{The expected revisit time over different work distance and target orbits} 
	\label{Fig.surf1} 
\end{figure}

Fig. \ref{Fig.surf1} and Table \ref{tab6} both show that a higher work distance can produce a V-shaped dependence of the $T_r$ curve for different targets. As the work distance $w$ increases, the minimum of this V-shape shifts to higher orbits. The results also suggest the existence of an optimal $w$ for each class of target orbit. At these optimized $w$, $T_2$ is relatively low compared with neighboring regions.

\subsection{Inclinations}
This subsection discusses how the inclinations of both the constellation and $U$ influence $T$.
Take $w=3000$ (m) and Elliptical orbit targets $U_2$, AA $r_2-h_0=2000$ m, PA $r_1=h_1=2000$ m, $i=s\frac{\pi}{12}$ and $i_U=t\frac{\pi}{12}$ where $t=0,1,2,3,4,5$, $s=1,2,3,4,5,6$.
\begin{table}[htbp]
\caption{$T$ vs. $i$ and $i_U$}
\centering
\resizebox{0.8\columnwidth}{!}{%
\begin{tabular}{c|c|c|c|c|c} 
 \toprule
 {$i_U/i$ } & {$\frac{\pi}{12}$} & {$\frac{2\pi}{12}$}&{$\frac{3\pi}{12}$}&{$\frac{4\pi}{12}$}&{$\frac{5\pi}{12}$}\\
 \midrule
 
 {$\frac{\pi}{12}$} & 2.583 & 2.354&2.121& 1.905&1.701 \\
 \hline
 {$\frac{2\pi}{12}$} & 2.354 & 2.241 &2.055&1.856&1.659\\
 \hline
 {$\frac{3\pi}{12}$} & 2.121 & 2.055 &1.939&1.772&1.586\\
 \hline
 {$\frac{4\pi}{12} $}& 1.905 & 1.856 &1.772&1.647&1.477\\
 \hline
 {$\frac{5\pi}{12} $}& 1.701 & 1.659 &1.586&1.477&1.322\\
 \bottomrule
\end{tabular}
\label{tab7}
}
\end{table}
Table \ref{tab7} shows the symmetry about the diagonal, indicating that the two inclinations are interchangeable.

Table (\ref{tab2}) sketches a simple result that as the inclination increases, the revisit time is decreased. Table \ref{tab7} shows additional information where if the inclinations of the targets and the constellation are $i$ and $i_U$, then we have:
\begin{equation}
    T_r(i,i_U)=T_r(i_U,i)
    \label{eq_ii}
\end{equation}
where $T_r$ decreases as both $i$ and $i_U$ increases, which is consistent with (\ref{tab2}).

\subsection{Analysis on how revisit time changes as orbital decays}

We first introduce the orbital decay $-\frac{da}{dt}$ from \eqref{eq_da}. Here, 
$a$ the semi-major axis is $h_0$.

\begin{figure}[H] 
	\centering 
	\includegraphics[width=0.45\textwidth]{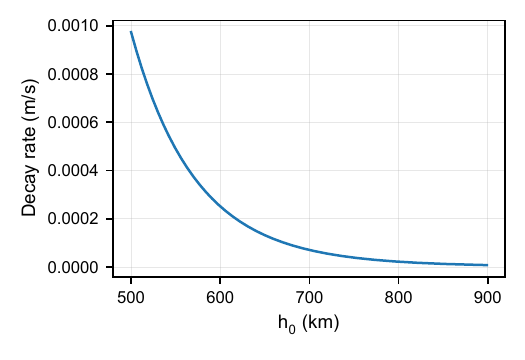} 
	\caption{The decay rate versus altitude $h_0$ with ballistic parameter=100 $kg/m^2$} 
	\label{Fig.main1} 
\end{figure}
Fig. \ref{Fig.main1} shows the relationship between decay rate and its altitude. It is shown that the orbital decay matters mostly at lower orbits. 

The orbital decay influences the geometric parameters of the constellation, which influences $T$. To show how orbital decay could change $T$, we can use $-\frac{dT}{dh_0}$. In the scenario of target $U_1$ , $w=3000$ m, $\Delta r=2000$ m, where $U_1$ is currently at $2000$ m above the constellation shell $U_{11}$, the position of the target is not dependent on $h_0$. The displacement of the orbit, or the change in $h_0$, must not exceed the region of detection defined by $w$ so that the target is observed. 
\begin{table}[htbp]
\caption{Derivative-1 $\frac{dT}{dh_0}$ vs. $h_0$ ,$\delta h=0.01$km}
\begin{center}
\begin{tabular}{c c} 
 \toprule
 $h_0$$\angle$ & 0.8 \\ 
 \midrule
 549.5 (km) & 0.49 (day/km)\\
 \hline
 550.0 (km)& 0.76 (day/km)\\
 \hline
 550.5 (km) & 1.35 (day/km)\\
 \hline
 551.0 (km) & 3.03 (day/km)\\
 \hline
 551.5 (km) & 12.1 (day/km)\\
 \bottomrule
\end{tabular}
\label{tab8}
\end{center}
\end{table}

For a target that is always $2000$ m above the constellation shell $U_{12}$, the position of the target is dependent on $h_0$. Parameters $h_0\in[500,2000]$ km uniformly distributed are chosen to cover the whole feasible region and show the trend.
\begin{table}[htbp]
\caption{Derivative-2 $\frac{dT}{dh_0}$ vs. $h_0$ ,$\delta h=10$km}
\begin{center}
\begin{tabular}{c c} 
 \toprule
 $h_0$$\angle$ & 0.8 \\ 
 \midrule
 500 (km) & 0.00054 (day/km) \\
 \hline
 1000 (km)& 0.00060 (day/km)  \\
 \hline
 1500 (km)& 0.00066 (day/km) \\
 \hline
 2000 (km)& 0.00073 (day/km) \\ 
 \bottomrule
\end{tabular}
\label{tab9}
\end{center}
\end{table}

For details, Table \ref{tab8} is  $\frac {\partial T(U(h_0),S(h_0))}{\partial h_0}$, and Table \ref{tab9} is  $\frac{dT(U(h_0),S(h_0))}{d h_0}$where $T(U,S)$ explicitly depends on $h_0$ by having $S$ as the geometrical parameters of the satellite constellation.

$\frac{dT_r}{da}$ stays positive in Tables \ref{tab8} and \ref{tab9}, suggesting that under the same magnitude, a low-orbit constellation is always more effective. This suggests that the orbital decay, if it does not influence $C_a$, is generally beneficial to the performance of this SSA system.

\subsection{Relationship between $\alpha$, $w$, and $T_r$}
In this subsection, we discuss the relationship between pointing angle $\alpha$, work distance$w$, and the expected revisit time$T_r$. Set target inclination to be $0$ and the constellation inclination to be $\frac{\pi}{2}$. All the values are taken between $w\in[0,100]$ km and $\alpha \in [0,90^{\circ}]$. As aforementioned in the subsection on work distance, a typical extreme of $2k\times 2k$ resolution optical sensors is around $w=100$ km, the choice of $w$ is justified.

\begin{figure}[htp]  \vspace{-10pt}
	\centering 
	\includegraphics[width=0.5\textwidth]{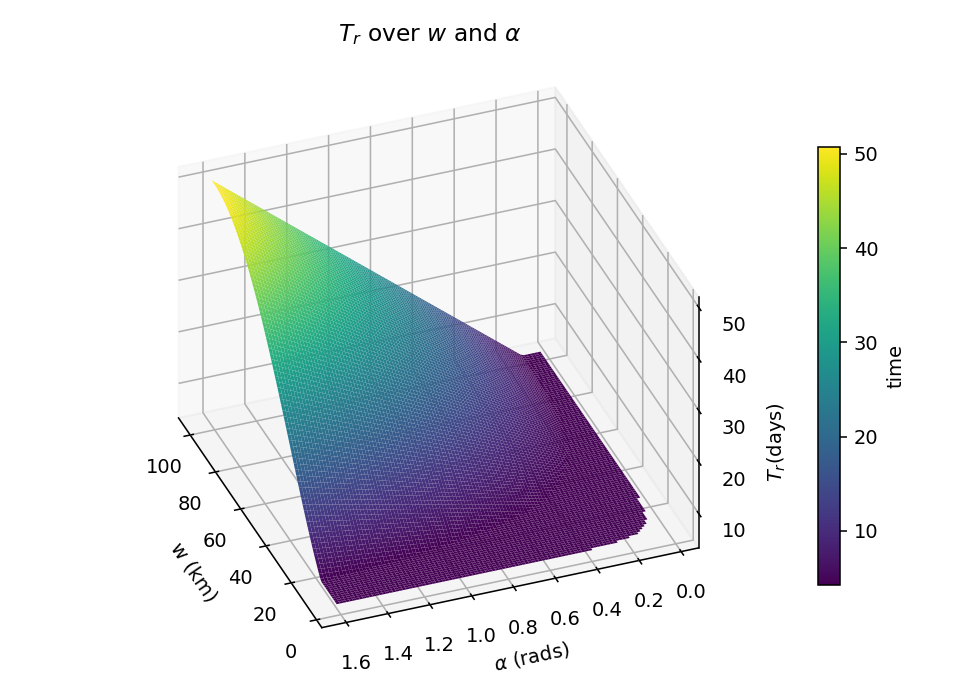} 
	\caption{The expected revisit time over different work distance and pointing angle at circular target orbit} 
	\label{Fig.surf2}  \vspace{-8pt}
\end{figure}

Fig.\ref{Fig.surf2} illustrates how the revisit time varies with $w$ and pointing angle for a given circular target orbit. The surface indicates that smaller pointing angles and shorter working distances can significantly improve revisit performance.

\begin{figure}[htp] 
	\centering 
	\includegraphics[width=0.5\textwidth]{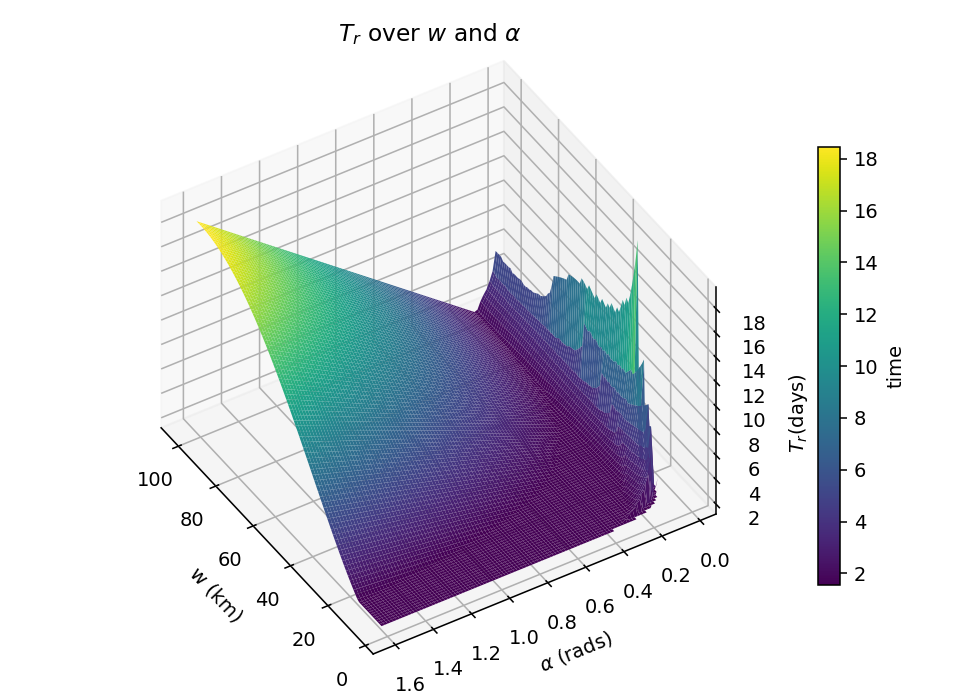} 
	\caption{The expected revisit time over different work distance and pointing angle at $s=9$ target orbit} 
	\label{Fig.surf3} 
\end{figure}

Fig.\ref{Fig.surf3} illustrates the circumstance when $s=9$, which makes the elliptic orbit mostly outside the range of detection. This is a representative circumstance for most targets. We noticed that there is a soft peak at $\alpha=\frac{\pi}{2}$ and high work distance, but the values near $\alpha=0$ is changed greatly compared with Fig. \ref{Fig.surf2}. There exists an optimal range of pointing angle $\alpha$, similar to the behavior observed in Fig. \ref{Fig.surf1}.

\begin{figure}[H]  \vspace{-12pt}
	\centering 
	\includegraphics[width=0.5\textwidth]{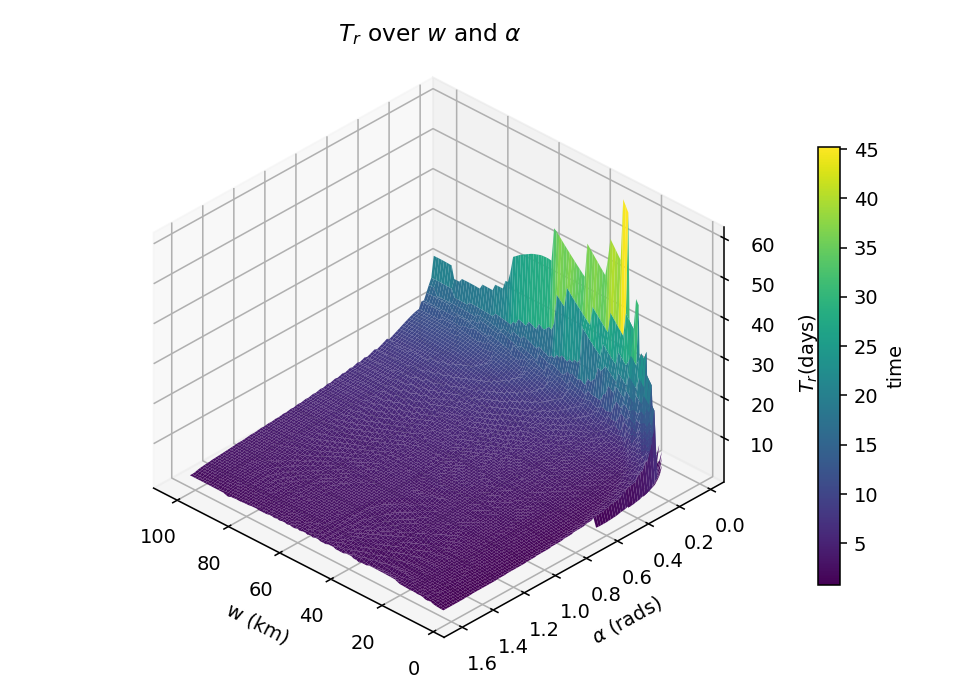} 
	\caption{The expected revisit time over different work distance and pointing angle at $s=100$ target orbit perspective 1} 
	\label{Fig.surf42} \vspace{-18pt}
\end{figure}

\begin{figure}[htp] 
	\centering 
	\includegraphics[width=0.5\textwidth]{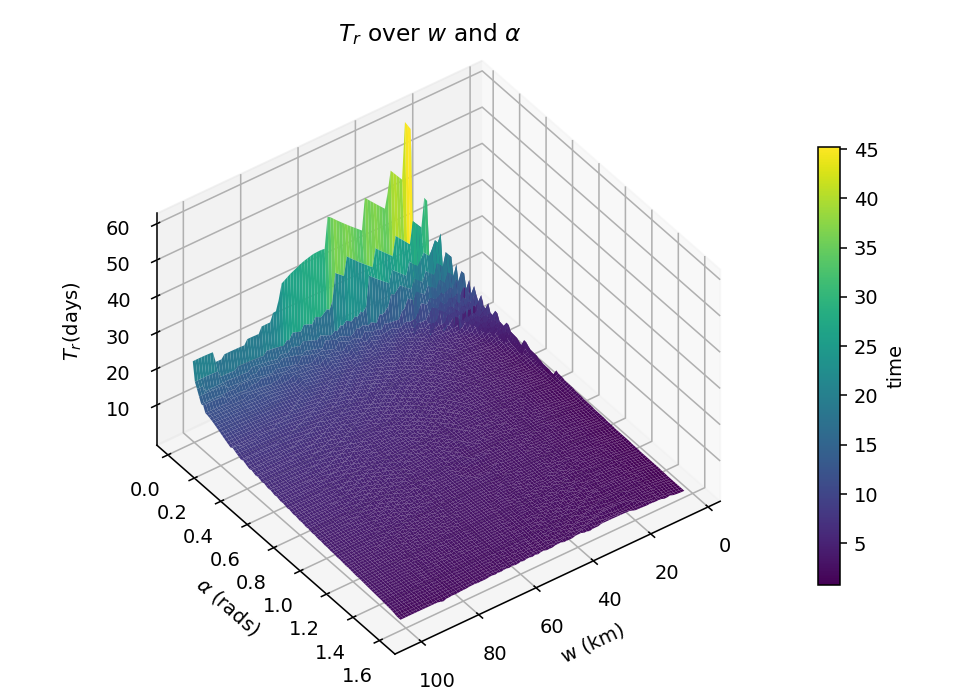} 
	\caption{The expected revisit time over different work distance and pointing angle at $s=100$ target orbit perspective 2} 
	\label{Fig.surf41}  \vspace{-10pt}
\end{figure}

In Fig. \ref{Fig.surf42}, the original viewing angle is retained, whereas in Fig. \ref{Fig.surf41}, the perspective is changed. Both figures display the same surface for $s=100$. The previous peak at $\alpha=\frac{\pi}{2}$  and high work distance no longer exists; instead, it is replaced by a relatively flat basin-like bottom. Significantly, a larger pointing angle is usually better for high orbits.

Taking the above results together, we conclude that, for a given class of target orbits, there exists an optimal combination of $w$ and pointing angle. This conclusion is obtained under the assumption of constant optical sensor resolution. Moreover, in most cases, there also exists an acceptable range of $\alpha$ and $w$ that is close to the optimized solution for each case.

\subsection{Result Discussion}

%

The numerical results are in good agreement with the theoretical analysis. In particular, the expected scaling behavior of $T$ with respect to the $t$ is clearly observed. 

%
The results highlight the strong sensitivity of revisit performance to key geometrical parameters. The inclination $i$, work distance $w$, and pointing angle $\alpha$ all have significant and related effects. 

Several non-intuitive behaviors are also observed. For example, increasing $p$ does not always improve performance because both too large and too small $p$ can result in under-utilization of available satellites. The interchangeability between $i$ and $i_U$ is also discovered.

\section{Conclusion}
This paper presents a geometrical analytical model framework for estimating $T_r$ and other metrics in SSA tasks. The proposed framework carries several important implications, and the defined objectives were successfully met, yielding the following key insights.

The proposed model is well-suited for SSA tasks involving optical sensors, while remaining broadly applicable to a wide range of scenarios. Its closed-form analytical derivation ensures transparency and interpretability, making it highly adaptable to varying conditions and easily modified to meet further requirements. Moreover, the model extends beyond existing closed-form approaches by incorporating more general constellation configurations, target orbits, and observation geometries.

Multiple metrics and parameter sensitivities are identified and illustrated. Several characteristic time schemes are also proposed and evaluated. From a practical perspective, the results suggest that effective constellation design requires a balance between satellite distribution and observation parameters. The proposed model is capable of large-scale parameter exploration and scanning to find other relationships.  

Our future work will build on the additional findings arising from this study. We noticed that the $C_a$ is strongly influenced by the number of orbital layers in the constellation. For constellations on elliptical orbits, however, it remains possible to trade situational-awareness efficiency evaluated by $T$ and $T_r$ against $C_a$. Even with a relatively small number of satellites, the system can still provide long-term awareness capability for space targets over a broad range of orbital regions and different targets.

By using variations of the proposed model as objective functions, recent studies \cite{XUE2024} have shown that reinforcement-learning-based algorithms can reduce computational cost. 

We also observe from \eqref{eq_Tra} that target-orbit characteristics significantly affect the evaluation of system performance. In practice, space targets are not fully random because of orbital decay and other dynamical constraints, but are more naturally distributed in structured groups. Therefore, a simple Monte Carlo treatment over all possible targets may not be the most appropriate way to assess overall system performance.

\section*{Acknowledgment}

We acknowledge the support of the Natural Sciences and Engineering Research Council of Canada (NSERC), [funding reference number RGPIN-2022-03364], and Research Manitoba.

\bibliographystyle{IEEEtran}
\bibliography{references}

\end{document}